\documentclass[final,3p,times,twocolumn]{elsarticle}

\usepackage{lineno}
\usepackage{graphics}
\usepackage{graphicx}
\usepackage{amssymb}
\usepackage{amsthm}
\usepackage{amsmath}
\usepackage{upgreek}
\usepackage{subfigure}
\usepackage{wasysym}
\usepackage{textcomp}
\usepackage{lineno}
\usepackage{booktabs}
\usepackage{paralist}

\biboptions{square}
\journal {Nuclear Instruments and Methods in Physics Research Section A}

\begin{document}
\begin{frontmatter}

\title{Studies on ion back-flow of Time Projection Chamber based on GEM and anode wire grid}

\cortext[cauthor]{
E-mail: xuqh@sdu.edu.cn}

\author[1:sdu]{Shuai Wang}
\author[1:sdu]{Fuwang Shen}
\author[1:sdu]{Xiao Zhao}
\author[2:sdu]{Zhihang Zhu}
\author[1:sdu]{Fangang Kong}
\author[1:sdu]{Changyu Li}
\author[1:sdu]{Qinghua Xu\corref{cauthor}}
\author[1:sdu]{Zhangbu Xu}
\author[1:sdu]{Chi Yang}
\author[1:sdu]{Chengguang Zhu}

\address[1:sdu]{Key Laboratory of Particle Physics and Particle Irradiation (MOE),
 Institute of Frontier and Interdisciplinary Science,
Shandong University, Qingdao, Shandong 266237, China}
\address[2:sdu]{
 School of Physics,
Shandong University, Ji'nan, Shandong 250100, China}

\begin{abstract}
Gated wires are widely used in Time Projection Chamber (TPC) to avoid ion back-flow (IBF) in the drift volume.
The anode wires can provide stable gain at high voltage with a long lifetime.
However, switching on and off the gated grid (GG) leads to a dead time and also limit the readout efficiency of the TPC.
Gas Electron Multiplier (GEM) foil provides a possibility of continuous readout for TPC, which can suppress IBF efficiently while keeping stable gain.
A prototype chamber including two layers of GEM foils and anode wires has been built to combine both advantages from GEM and anode wire.
Using Garfield++ and the finite element analysis (FEA) method, simulations of the transmission processes of electrons and ions are performed and results on absorption ratio of ions, gain and IBF ratio are obtained.
The optimized parameters from simulation are then applied to the prototype chamber to test the IBF and other performances.
Both GEM foils are run at low voltage (255V), while most of the gain is provided by the anode wire.
The measurement shows that the IBF ratio can be suppressed to $\sim$0.58$\%$ with double-layer GEM foils (staggered) at an effective gain about 2500 with an energy resolution about 10$\%$.
\end{abstract}
\begin{keyword}
TPC; GEM; IBF; Garfield++; MWPC
\end{keyword}

\end{frontmatter}
\section{Introduction}

In traditional Time Projection Chamber (TPC) \cite{ref:detector_01} with Multi Wire Proportional Chamber (MWPC) as read out, gated grid (GG) is used to prevent space charge distortion caused by ion back-flow (IBF) which could affect the drift and diffusion of ionized electrons and degrade performance of the detector.
GG is usually controlled by applying pulse voltage to the gated wires. When the gate is closed, the ions are blocked from flowing back to the drift volume.
It must remain closed until the ions generated in avalanches by the anode wires drift to the gated wires and are all absorbed.
This mechanism will lead to a dead time and thus limit the readout efficiency.

In recent years, the rapid development of Gas Electron Multiplier (GEM) \cite{ref:detector_02,ref:detector_03}, has made it possible to effectively suppress the ion back-flow (IBF) and minimize the dead time of the detector with continuous readout.
Therefore, GEM becomes a good choice to serve as a more efficient readout of TPC detector, for example, in the proposed TPC upgrade at ALICE experiment  \cite{ref:detector_x5,ref:detector_04}.
In addition, the combination of Micromegas and GEM has also been verified to be a feasible choice in several experiments \cite{ref:detector_05}.

The advantage of continuous readout is obvious for high rate experiment, but is also subject to space charge issue due to ion back flow. 
We need to keep the space charge density in an acceptable range while the IBF shall be controlled as low as possible. 
In this paper, we propose a new structure of TPC consisting of both anode wires and GEMs, intended for experiment environment like future EIC collider, where collision rate is high, but the multiplicity is relatively low.
The anode wires will provide the most of the effective gain, which has been found quite stable in TPC for example at STAR experiment for successful running for about 20 years without sign of aging.
Then GEM foils will replace the gated wires, to suppress the IBF without providing significant part of the effective gain by running at low voltage ($<$300 V), which helps to reduce the probability of discharge.
As will be discussed further with detail in Section 4.3, the charge density with $e+p$ collision at future EIC collider will be comparable to that at RHIC-STAR experiment. Therefore the proposed TPC structure consisting anode wire and GEM foils provides a possible way of realizing continuous readout scheme for high luminosity but low multiplicity experiment as EIC collider.

In order to study the combination of GEM and anode wire, a prototype TPC chamber which consists of two layers of GEM foils and anode wires was built.
Compared with traditional MWPC chamber, the gated and shield wires are removed while the anode wires are kept, then two layers of GEM foils are placed above the anode wire grid.
The authors gained valuable experiences by completing the MWPC assembly for STAR inner TPC upgrade,  so some of the key parameters related with wires here are similar as STAR TPC\cite{ref:detector_07}, for example the gas, drift field, read out electronics etc.
The detailed experimental setup, simulation results using Garfield++, and measurements of key performance parameters in particular IBF ratio with the prototype will be discussed in the following sections.

\section{Experimental setup}

\begin{figure}[htbp]
\begin{center}
\centerline{\includegraphics[width=8.0cm]{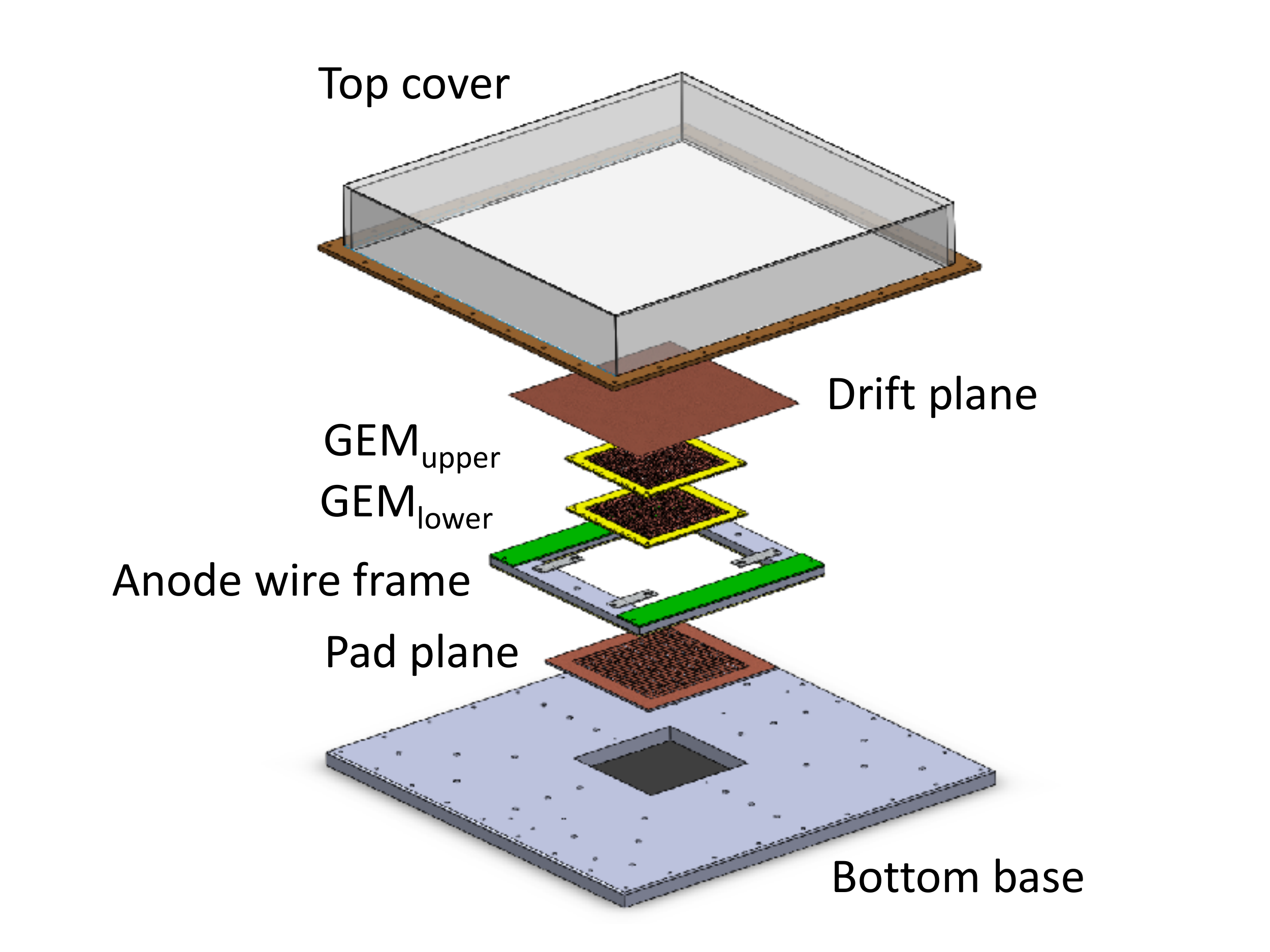}}
\caption{Exploded 3D view of the prototype chamber.
\label{fig:3D view of the detector}}
\end{center}
\end{figure}

The structure of the prototype chamber is shown in Figure~\ref{fig:3D view of the detector}, which consists of a top cover, drift plane, two layers of GEM foils, anode wire frame, pad plane and bottom base. Its overall size is 45.5cm $\times$45.5cm$\times$6.5cm.
Above the bottom base, there is a 10cm$\times$10cm pad plane with the pad size of 15.5mm$\times$5.5mm and 0.5mm gap between adjacent pads.
The anode wire frame consists of 28 anode wires with a pitch of 4mm.
20$\upmu$m diameter gold-plated tungsten wires with a tension of 0.5 Newton were used \cite{ref:wangxu}.
Then two 10cm$\times$10cm standard GEM foils \cite{ref:Yang15} were mounted above the anode plane, with a pitch of 140$\upmu$m, 70$\upmu$m outside diameter, 50$\upmu$m internal diameter, 50$\upmu$m kapton thickness, and 5$\upmu$m copper thickness on both sides.
A 10cm$\times$10cm copper sheet with thickness of 50$\upmu$m is used as the drift plane.
The top cover is made of plexiglass covered with a layer of copper to prevent electromagnetic noise.

P10 ($90$$\%$Ar+$10$$\%$CH$_4$) with 0.1$\%$ mixing precision is used as the working gas, with the same choice as STAR TPC\cite{ref:detector_07}.
A differential pressure transmitter (Huba 699) with an accuracy of 0.02mbar is used to control the relative pressure inside the test chamber.
The temperature monitoring is realized by a Arduino-BMP180 module with an accuracy of 0.5$^\circ$C.
After filling in the P10 gas, the temperature and pressure were controlled to 25$^\circ$C (by air-conditioning) and +2mbar above the atmospheric pressure.
During one typical measurement cycle, the variance of the temperature is less than 0.8\%, and the variance of the pressure is less than 0.1\%.

\begin{figure}[htbp]
\begin{center}
\centerline{\includegraphics[width=8.0cm]{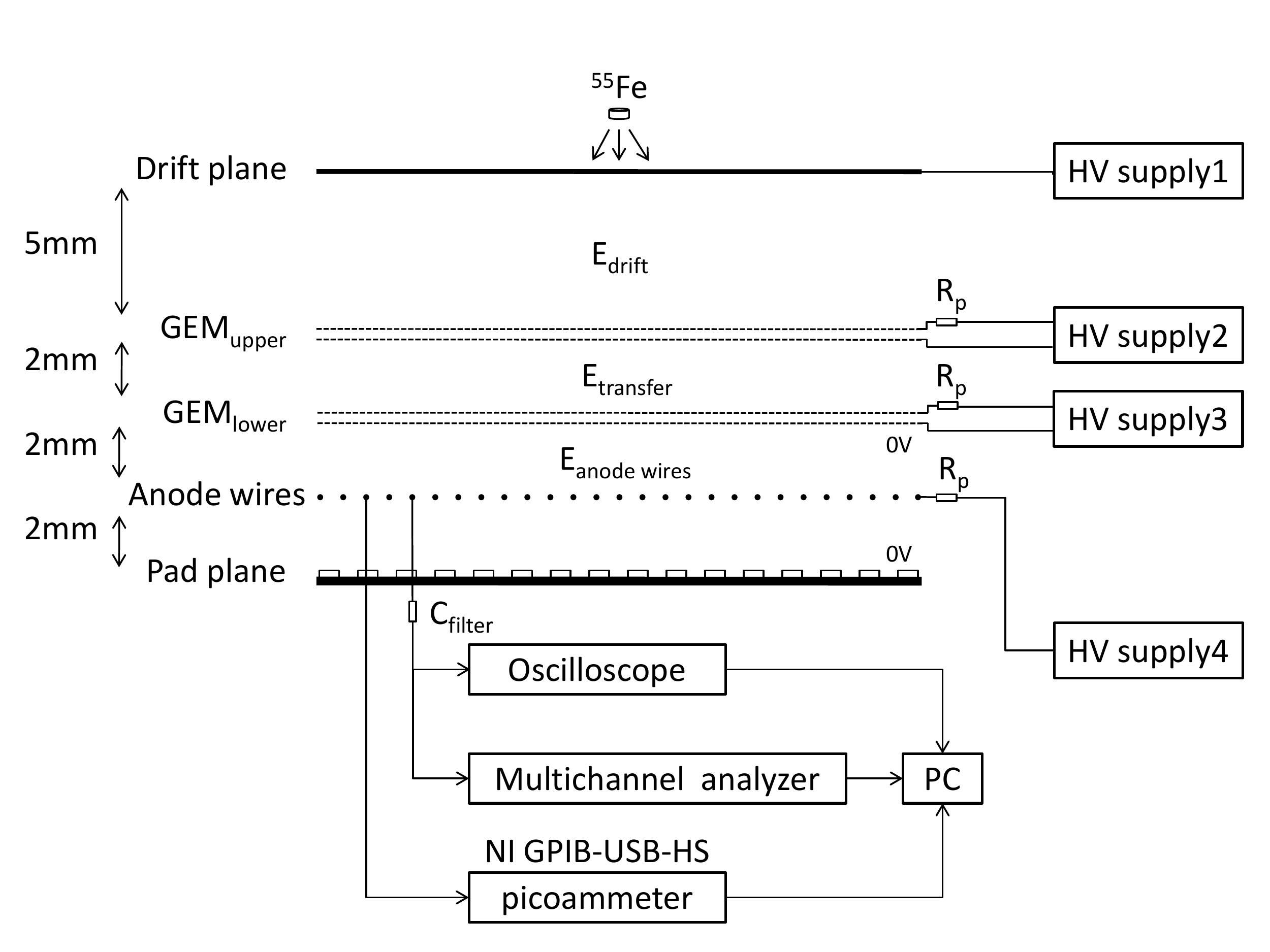}}
\caption{The layout of the chamber and the test system based on $^5$$^5$Fe X-ray. $E_{\rm drift}$, $E_{\rm transfer}$ and $E_{\rm anode\;wires}$ are the E-field of the drift volume, E-field between two GEM foils and
E-field between GEM$_{\rm lower}$ and the anode wires respectively.
\label{fig:Sketch map of the detector}}
\end{center}
\end{figure}

The distances between each plane are shown in Figure~\ref{fig:Sketch map of the detector}, which also describes the details of the testing system.
To test the performance of the prototype, an $^5$$^5$Fe source with a diameter of 5mm was used to measure the gas gain, energy resolution and IBF.
Several protective resistors R$_{\rm p}$ (10M$\Omega$) were used to protect the GEM foils \cite{ref:detector_x3} , and capacitor (1.3nF) was used to filter the signal for readout.

Oscilloscope (Tektronix MDO4054B-3) was used to capture the pulse signal on wire or pad with a high gain (more than $\sim$10000) using LabVIEW.
For a low gain (less than $\sim$10000), Multi-Channel Analyzer (Amptek MCA 8000D) was used to digitize the pulse signal.
The current was measured using a picoammeter (Keithley 6482, with a resolution of 1fA).

\begin{figure}[htbp]
\begin{center}
\subfigure[]{
\includegraphics[width=7.2cm]{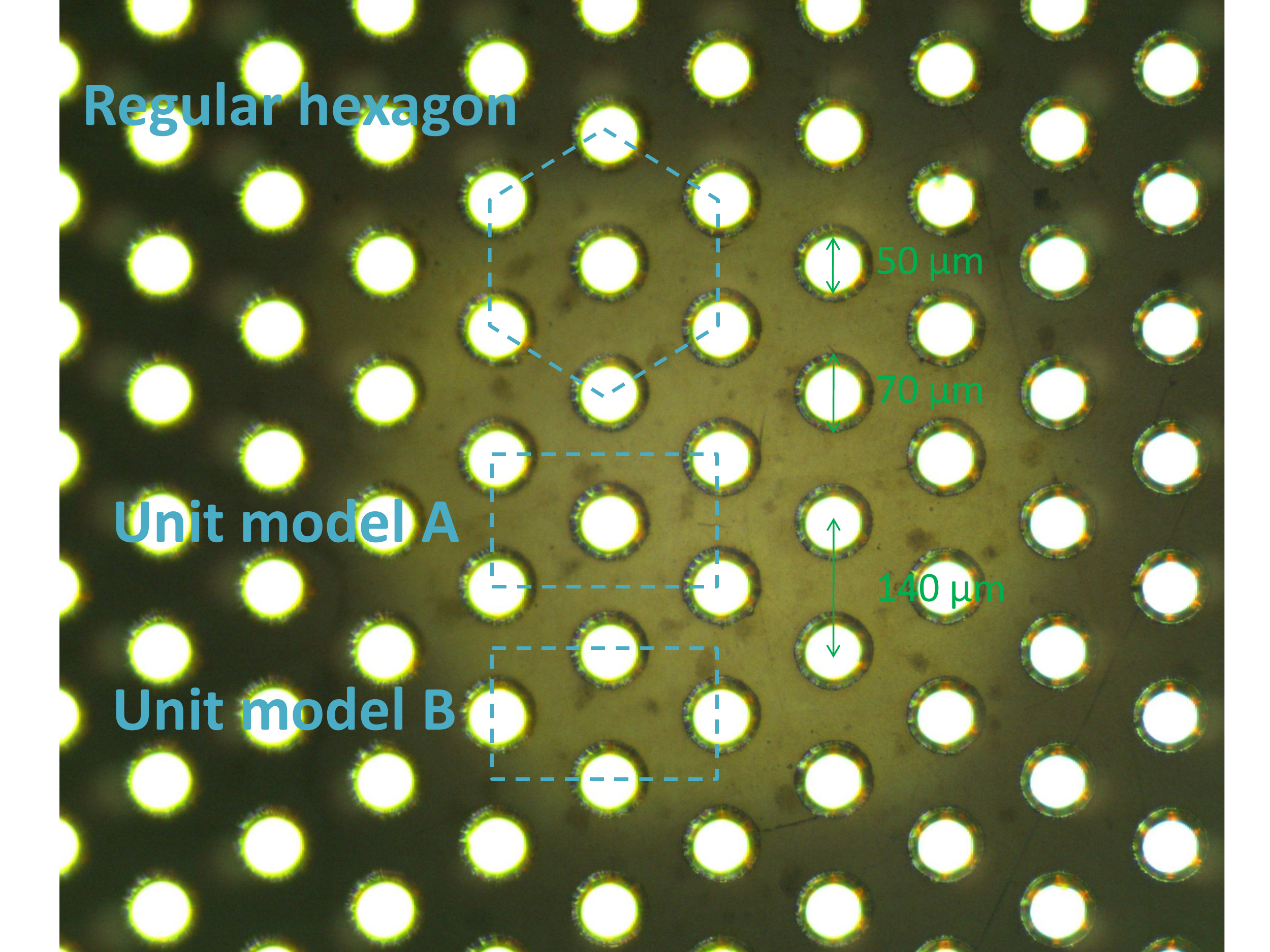}}
\subfigure[]{
\includegraphics[width=7.2cm]{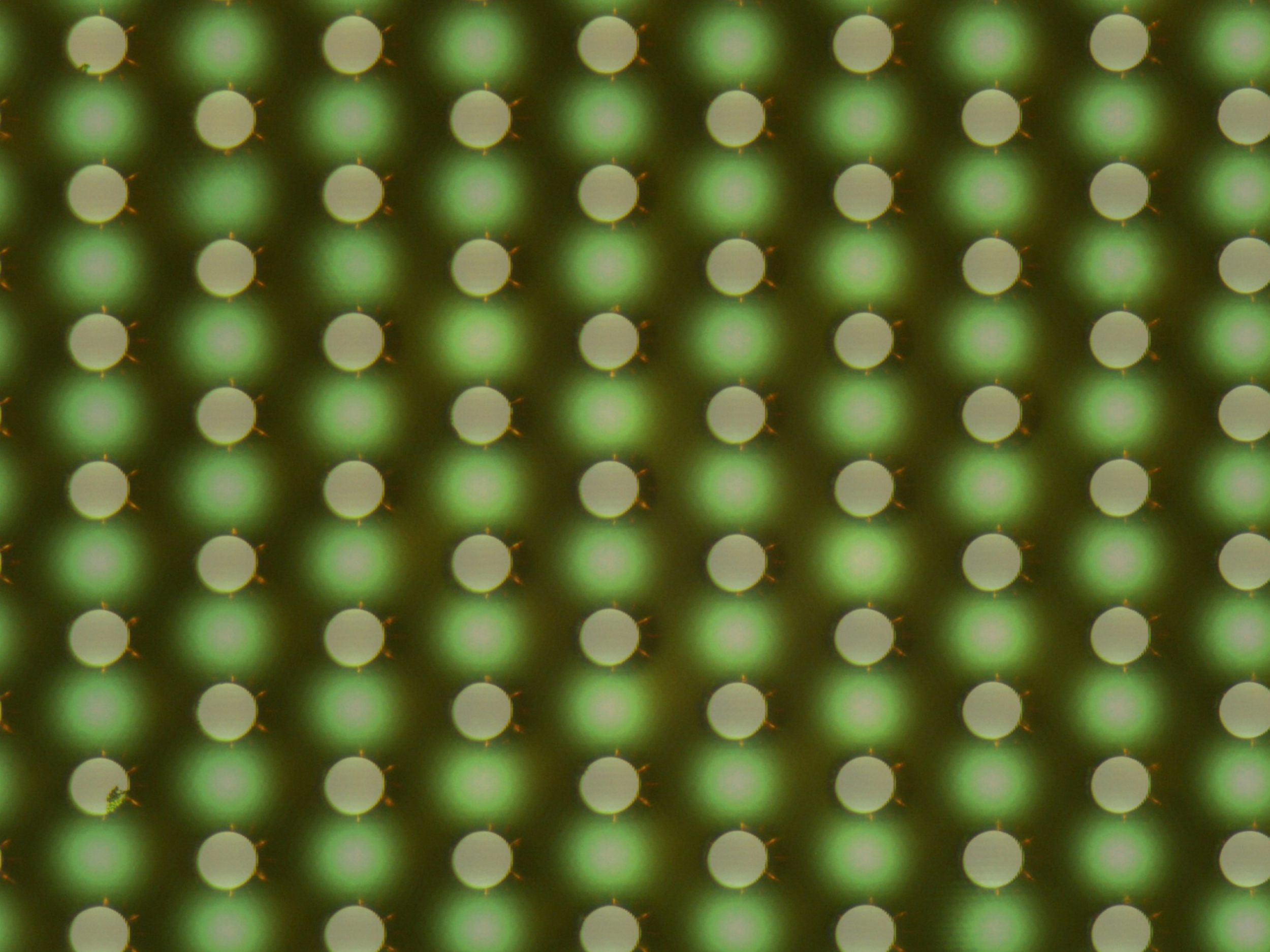}}
\subfigure[]{
\includegraphics[width=7.2cm]{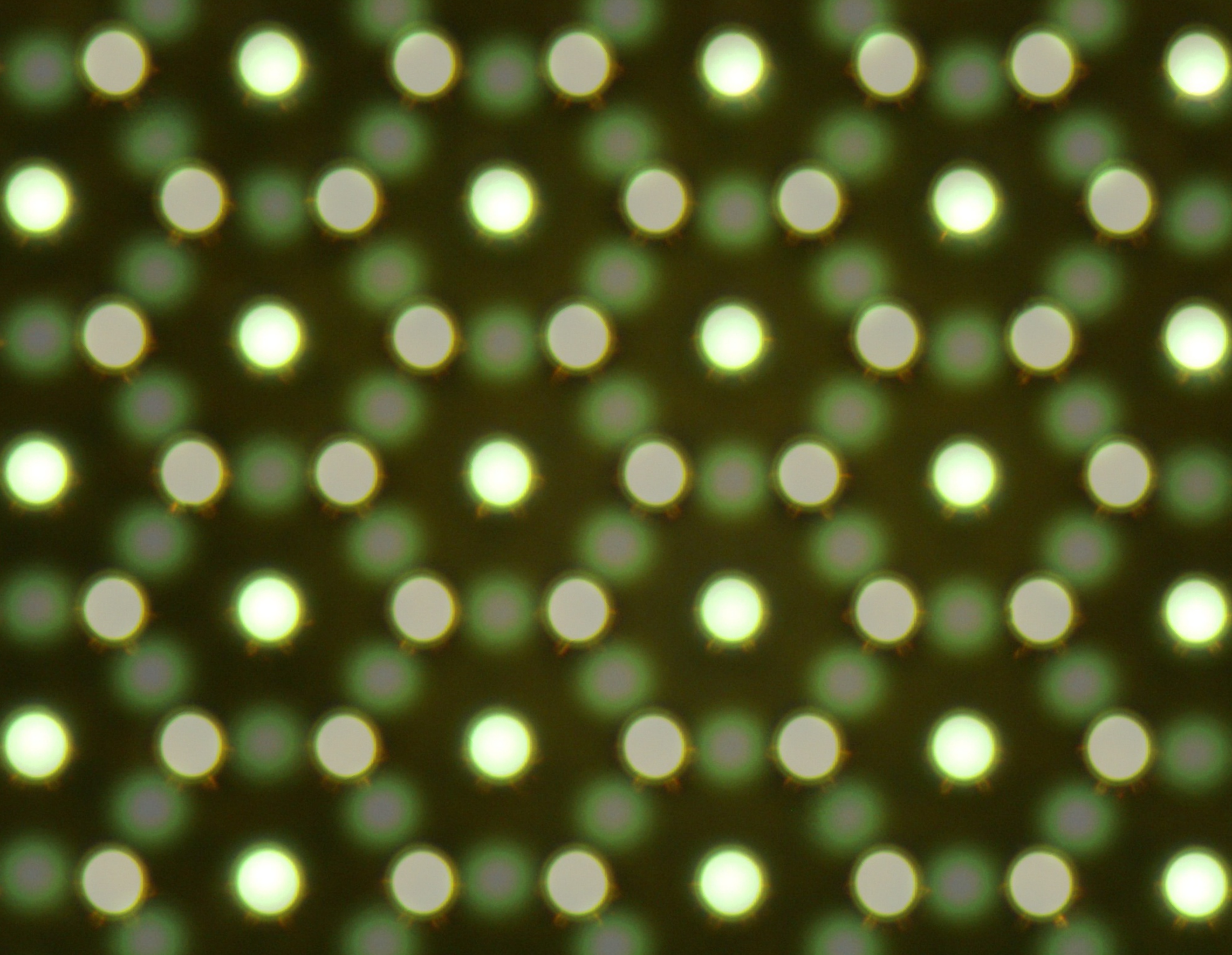}}
\caption{(a) Structure of GEM foil. (b) Double-layer GEMs with staggered mode realized by moving one GEM foil upward/downward by 70$\upmu$m relative the other one. The brighter spots are the holes on the upper foil and the blurred spots are the holes of the lower one. (c) Double-layer GEMs with staggered mode realized by rotating one GEM foils for 90$^\circ$ relative to the other one. The brightest spots are the aligned holes after rotation. The brighter spots are the holes of upper foil and the blurred spots are the holes of lower GEM foil.}
\label{fig:GEM study}
\end{center}
\end{figure}

Several ways of shielding have been applied to reduce the noise from possible interferences. The top cover is shielded with a copper sheet, which is grounded, same as the aluminum base. The picoammeter to measure the current is also put in a copper-covered box.
All the connection cables are equipped with anti-interference magnetic ring to reduce electromagnetic interference.

The structure of the GEM foils have been verified with microscope and is shown in Figure~\ref{fig:GEM study}(a).  As can be seen, the holes are distributed in a hexagon shape. 
The sizes of the GEM holes are not checked one by one, and the possible deviations may contribute to the performance difference in simulation and measurement shown later in Section 4 although this effect could be small.
Since the relative position of two GEM foils is quite relevant to the performance on the IBF, different choices are made both in simulation and experimental setup.
In the first choice, the two GEM foils are placed in exact alignment to each other in the sense that holes are along the same line in the direction of electric field, which is refereed as "non-staggered" mode later.
In the second choice, they are mounted with complete disalignment to each other for the relative position of holes, named by "staggered" mode.
As shown in Figure~\ref{fig:GEM study}(b) (taken by microscope), one ideal staggered mode can be realized by shifting one foil upward/downward by 70$\upmu$m relative to the other foil.
This ideal staggered mode as in Figure~\ref{fig:GEM study}(b) can be achieved in the simulation (will be discussed in next section), but it is a bit difficult in experiment which is subject to the mechanical precision while mounting the GEM foils.
In experiment, another staggered mode is realized by rotating one GEM foils for 90$^\circ$ relative to the other one, as shown in Figure~\ref{fig:GEM study}(c) (from microscope). 
Here the rotation relies on the square shape and the rotating axis needs to be in the foil center, but this close-to-ideal staggered mode is less affected by the mounting precision as in Figure~\ref{fig:GEM study}(b).
In summary,  the staggered mode as Figure~\ref{fig:GEM study}(b) will be used in simulation in Section 3, which can be served as guide on ideal IBF suppression. 
For the experimental testing with our prototype chamber in Section 4,  the staggered mode as Figure~\ref{fig:GEM study}(c) will be used and compared with simulation results.

The principle of voltage set up is that the anode wires provide most of the gain while the GEMs pre-amplify the signal so that GEMs can be operated at a relatively low voltage in order to reduce their discharge probability.
The high voltage on the anode wires was set to be $\sim$1000V in order to provide the most of the gain, and the anode wire gain is about 2400 at high voltage of 1120V, similar as STAR inner TPC \cite{ref:detector_07}.
The voltage for the GEM foils, $\Delta$$V_{\rm GEM}$ ($\Delta$$V_{\rm GEM_{\rm upper}}$=$\Delta$$V_{\rm GEM_{\rm lower}}$) was set to be lower than 300V, so the spark probability is significantly reduced.
Since the amplitude of discharge signal is much higher than that of ordinary signal, to observe the discharge on GEMs, we adjusted the oscilloscope trigger threshold to filter out the ordinary signals and to trigger on discharge signals with a much higher amplitude.
Then the signals over the threshold will be recorded.
$\Delta$$V_{\rm GEM}$=255V was tested and found to be stable without discharge for certain period of time.
In the follow studies, we fix the $\Delta$$V_{\rm GEM}$ at 255V for both GEM foils and controlled the gain by adjusting the voltage on anode wires (940-1160V).
For the effects from electric field ($E_{\rm drift}$ and $E_{\rm transfer}$) settings, simulations have been done to provide us guidance and will be further checked with experimental results.

\section{Simulation results}

The simulation of the prototype chamber based on GEMs plus anode wires was done with Garfield++ to provide guidance for experimental tests \cite{ref:detector_x2}.
The same chamber structure as in Figure~\ref{fig:3D view of the detector} has been realized in the simulation.
The finite element analysis (FEA) method is used in the modeling. The FEA method divides the chamber volume into tiny elements. Each element consists of several nodes where the potential is calculated.
The more meshes being divided, then more nodes were generated.  A rectangular geometry unit model was built to avoid the possibility of memory overload.

\begin{figure}[htbp]
\begin{center}
\subfigure[]{
\label{fig:Electric field1}
\includegraphics[width=7.5cm]{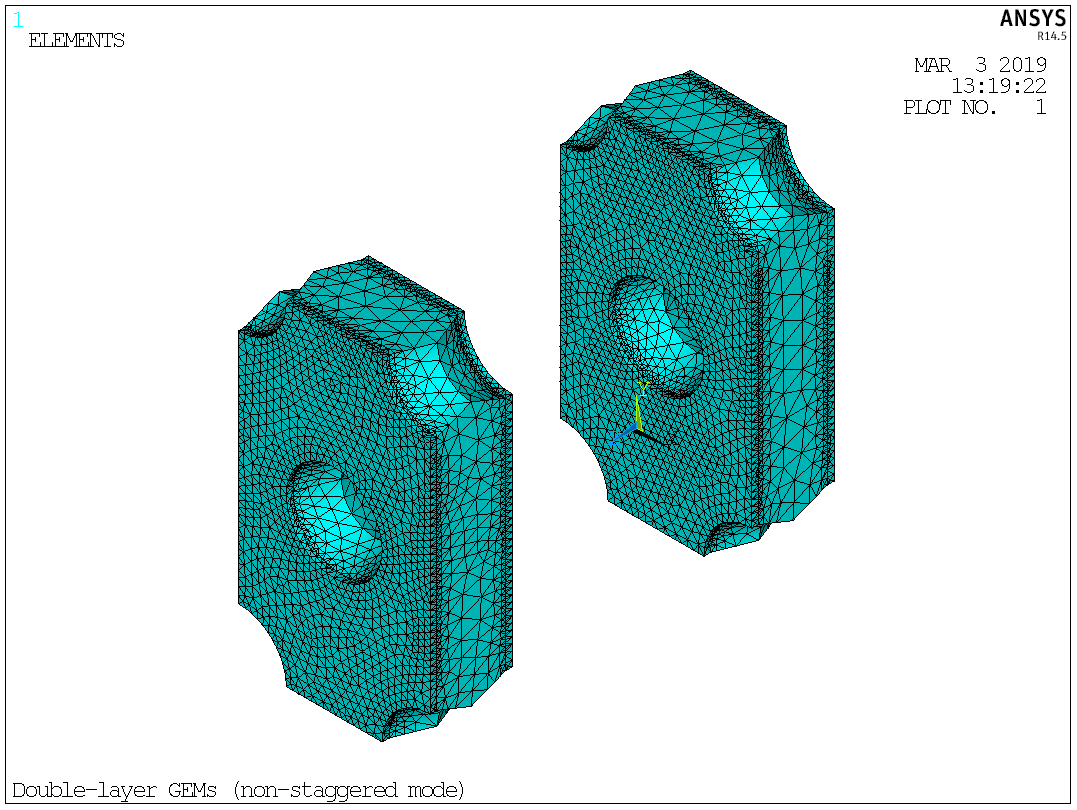}}
\subfigure[]{
\label{fig:Electric field2}
\includegraphics[width=7.5cm]{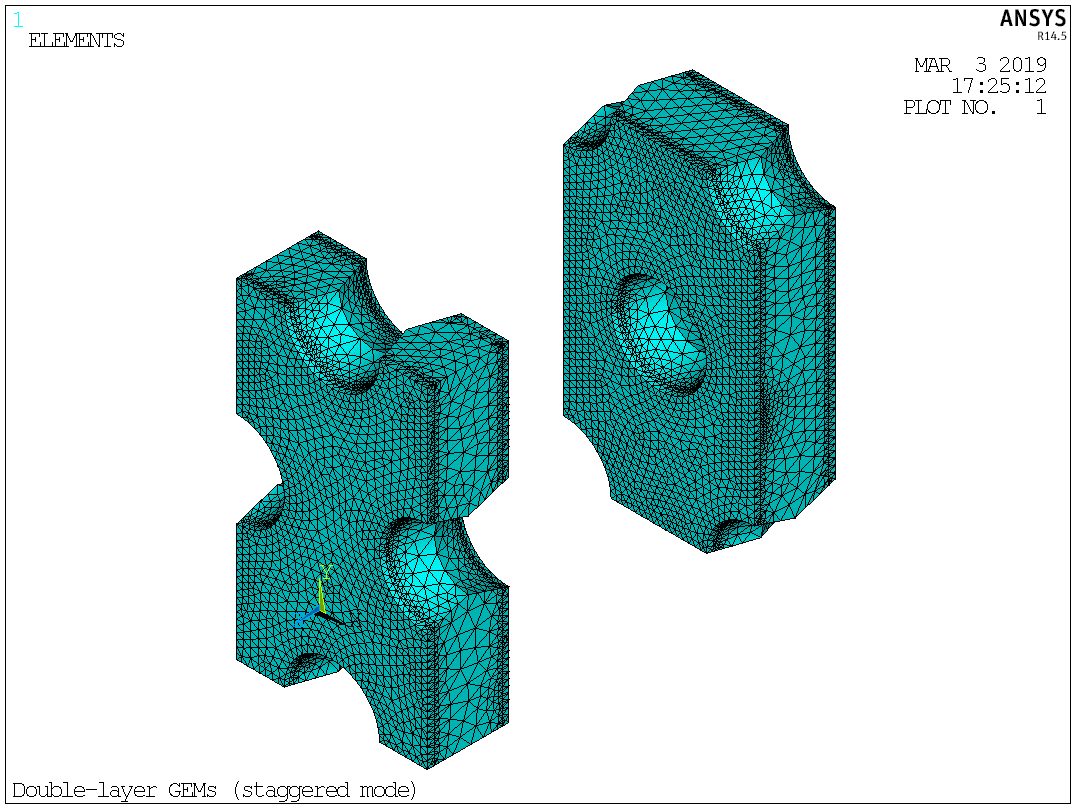}}
\caption{Visualizing the model of double-layer GEMs by ANSYS Parametric Design Language: (a) Non-staggered mode; (b) Staggered mode.}
\label{fig:Electric field}
\end{center}
\end{figure}

With ANSYS Parametric Design Language, unit model A and unit model B as shown in Figure~\ref{fig:GEM study}(a) with a size of 140.0$\upmu$m$\times$242.5$\upmu$m were built.
For the relative mode of two layers of GEM foils (staggered or non-staggered as discussed in previous section), two models were established
as shown in Figure~\ref{fig:Electric field}.
In particular, the staggered mode as in Figure~\ref{fig:GEM study}(b) is simulated with unit model A for one GEM foil and the unit model B for the other foil.
The two units A and B were put in the same coordinate system and aligned in the vertical direction along the potential, which form into a  cube unit.
Then various parameters such as material, dielectric constant and resistivity were added to each part of the model.
After mesh generation and solution, the potential information carried by the cube unit will be generated.
Then it was imported into Garfield++ which can  expand the cube unit with potential information to the actual size (10cm$\times$10cm) by the mirror copy \cite{ref:detector_09}.
In summary, the full staggered mode as in Figure~\ref{fig:GEM study}(b) is realized in simulation with steps mentioned above. However, we note that the staggered mode by rotation method in experiment as in Figure~\ref{fig:GEM study}(c) can not be realized with ANSYS simulation.

\begin{figure}[htbp]
\begin{center}
\centerline{\includegraphics[width=8.3cm]{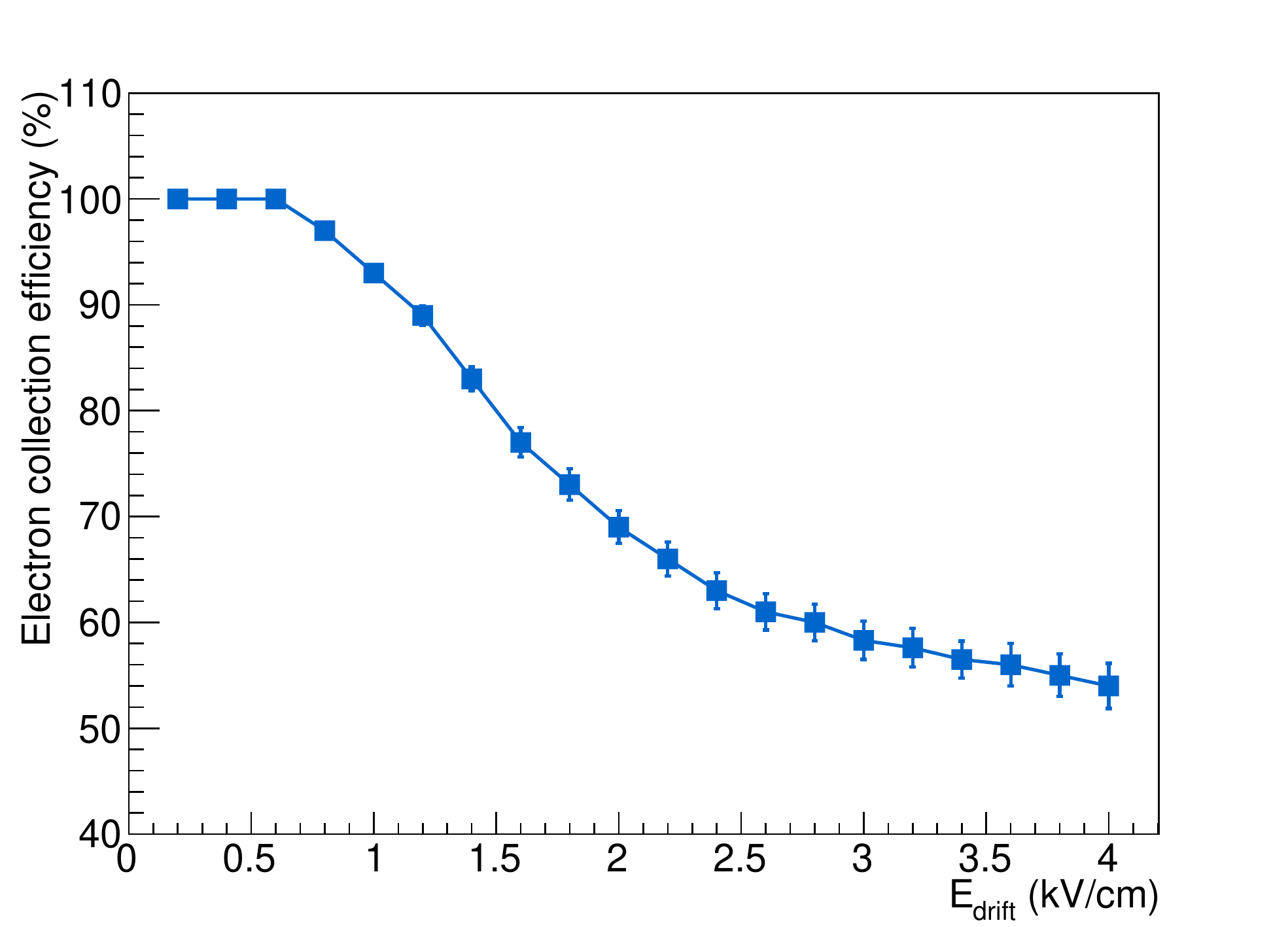}}
\caption{Simulation results of electron collection efficiency versus $E_{\rm drift}$.  Here $V_{\rm anode\;wires}$=1120V,  $\Delta$$V_{\rm GEM_{upper}}$=$\Delta$$V_{\rm GEM_{lower}}$=255V, $E_{\rm transfer}$=4.0kV/cm.
\label{fig:Drift-in efficiency of the electrons}}
\end{center}
\end{figure}

Simulation results of electron collection efficiency versus $E_{\rm drift}$ are shown in Figure~\ref{fig:Drift-in efficiency of the electrons}.
The electron collection efficiency is defined as the probability of electrons drifting into the GEM holes \cite{ref:detector_x1}.
The collection efficiency reaches about 100\% at low drift field and starts to decrease after 0.5kV/cm as expected. 

After the electrons enter the GEM holes, avalanches occur. 
The generated avalanche electrons will enter the next layer of GEM for amplification, and finally reach the surface of the anode wires where the last avalanches occur \cite{ref:detector_new01}. Each avalanche produces an equal amount of ions as electrons. These ions will flow back to the drift volume under the electric field. Most of the ions will be absorbed by GEM foils, and a small fraction of ions will reach the drift plane.  The IBF ratio, which is defined as the fraction of ions that flow back to the drift plane, is usually used to describe the ability to inhibit ions \cite{ref:detector_10}.
As mentioned earlier, the voltage for GEM here is chosen to run at low value with smaller gain about 4 for each layer at 255V,  to realize a smaller IBF and reasonable energy resolution. 

\begin{table}[htbp]
\begin{center}
\begin{tabular*}{7.8cm}{@{\extracolsep{\fill}}lll}
            \hline
            \;\;\;Position&Non-staggered&Staggered\\ \hline
            \;\;\;$\rm Drift\;plane$& (3.5$\pm$0.5)$\%$ & (0.4$\pm$0.1)$\%$
          \\\;\;\;$\rm U-copper_{top}$& (58.7$\pm$4.9)$\%$ & (44.4$\pm$4.1)$\%$
          \\\;\;\;$\rm U-kapton$& (1.9$\pm$0.2)$\%$ & (1.6$\pm$0.2)$\%$
          \\\;\;\;$\rm U-copper_{bottom}$& (1.6$\pm$0.2)$\%$ & (21.2$\pm$2.9)$\%$
          \\\;\;\;$\rm L-copper_{top}$& (0.3$\pm$0.1)$\%$ & (0.2$\pm$0.1)$\%$
          \\\;\;\;$\rm L-kapton$& (1.7$\pm$0.2)$\%$ & (1.5$\pm$0.2)$\%$
          \\\;\;\;$\rm L-copper_{bottom}$& (1.3$\pm$0.2)$\%$ & (1.2$\pm$0.2)$\%$
          \\\;\;\;$\rm Pad\;plane$& (31.0$\pm$3.4)$\%$ & (29.5$\pm$3.1)$\%$  \\ \hline
\end{tabular*}
\caption{Simulation for absorption fractions of ions by each part (staggered mode and non-staggered mode). $V_{\rm anode\;wires}$=1120V, $\Delta$$V_{\rm GEM_{upper}}$=$\Delta$$V_{\rm GEM_{lower}}$=255V, $E_{\rm drift}$=0.1kV/cm, $E_{\rm transfer}$=4.0kV/cm. U and L represent GEM$_{\rm upper}$ and GEM$_{\rm lower}$ respectively.}
\label{tab:1}
\end{center}
\end{table}

To simulate the IBF of the prototype, we first calculated the ion absorption fraction of each part under two modes of the two GEM foils (non-staggered and staggered mode respectively). Table~\ref{tab:1} shows the ion absorption fractions of each parts under $\Delta$$V_{\rm GEM}$=255V for both GEM foils, $E_{\rm drift}$=0.1kV/cm and $E_{\rm transfer}$=4.0kV/cm.
The uncertainties are obtained by repeating the simulation 500 times (RMS).
Most of the ions ($>$60$\%$) are absorbed by the upper layer GEM foil in both modes.
And $\sim$30$\%$ of ions flow into the pad plane, which is right below the anode wire grid.
The fraction of ions finally reaching the drift plane, effectively the IBF ratio, is (3.5$\pm$0.5)$\%$ and (0.4$\pm$0.1)$\%$ for non-staggered and staggered modes respectively.
Obviously, the staggered mode is more effective than the non-staggered mode to inhibit ions.

\begin{figure}[htbp]
\begin{center}
\centerline{\includegraphics[width=8.5cm]{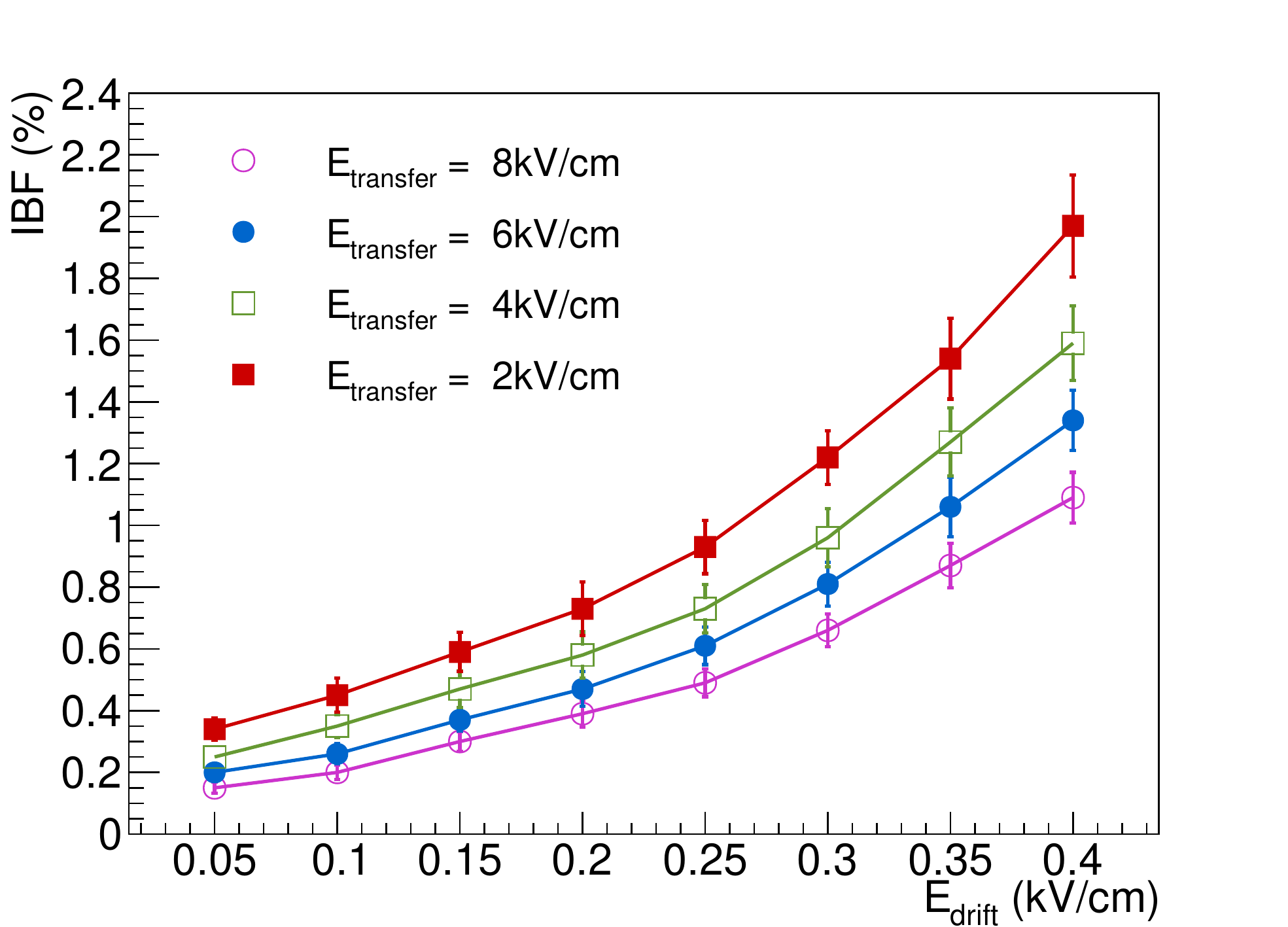}}
\caption{Simulation results of IBF ratio versus $E_{\rm drift}$ for several fixed values of $E_{\rm transfer}$ (2.0, 4.0, 6.0, 8.0kV/cm). Here $V_{\rm anode\;wires}$=1120V, $\Delta$$V_{\rm GEM_{\rm upper}}$=$\Delta$$V_{\rm GEM_{\rm lower}}$=255V.
\label{fig:E transfer and IBF}}
\end{center}
\end{figure}

Figure~\ref{fig:E transfer and IBF} shows the influence of $E_{\rm transfer}$ on IBF ratio versus $E_{\rm drift}$ under staggered mode for the two GEM foils.
For the same $E_{\rm drift}$, the IBF ratio decreases as the $E_{\rm transfer}$ increases.
When $E_{\rm drift}$=0.1kV/cm, $E_{\rm transfer}$$>$4.0kV/cm with anode wire voltage 1120V, IBF can be suppressed to less than 0.35$\%$.
It is worth noting that, for higher $E_{\rm transfer}$ ($>$3kV/cm), the avalanches begin to extend further into the transfer region resulting in a gain increase \cite{ref:detector_11}.
Therefore, we should avoid the application of high $E_{\rm transfer}$ configuration, and have to make a certain trade-off between suppressing ions flowing back and providing stable gain.
For this reason, $E_{\rm transfer}$ was set to 4.0kV/cm in the following tests.

\section{Experiment results}

In this section, testing results of the prototype chamber including effective gain, energy resolution and IBF ratio using $^5$$^5$Fe X-ray source will be provided.
$\Delta$$V_{\rm GEM}$=255V for both GEM foils, $E_{\rm transfer}$=4.0kV/cm and $E_{\rm drift}$=0.1kV/cm are set as default values.

\subsection{$^5$$^5$Fe spectrum and gain measurements}

The $^5$$^5$Fe source has been used for the testing, which mainly emits 5.9keV X-rays.
In P10 gas, it will produce approximately 225 electrons for the main peak ($\sim$5.9keV) or 110 electrons for escape peak ($\sim$2.9keV) \cite{ref:detector_x6}.
The avalanche on the anode wire will generate a pulse signal, which will be displayed on the oscilloscope and captured by LabVIEW.

\begin{figure}[htpb]
\begin{center}
\centerline{\includegraphics[width=8.5cm]{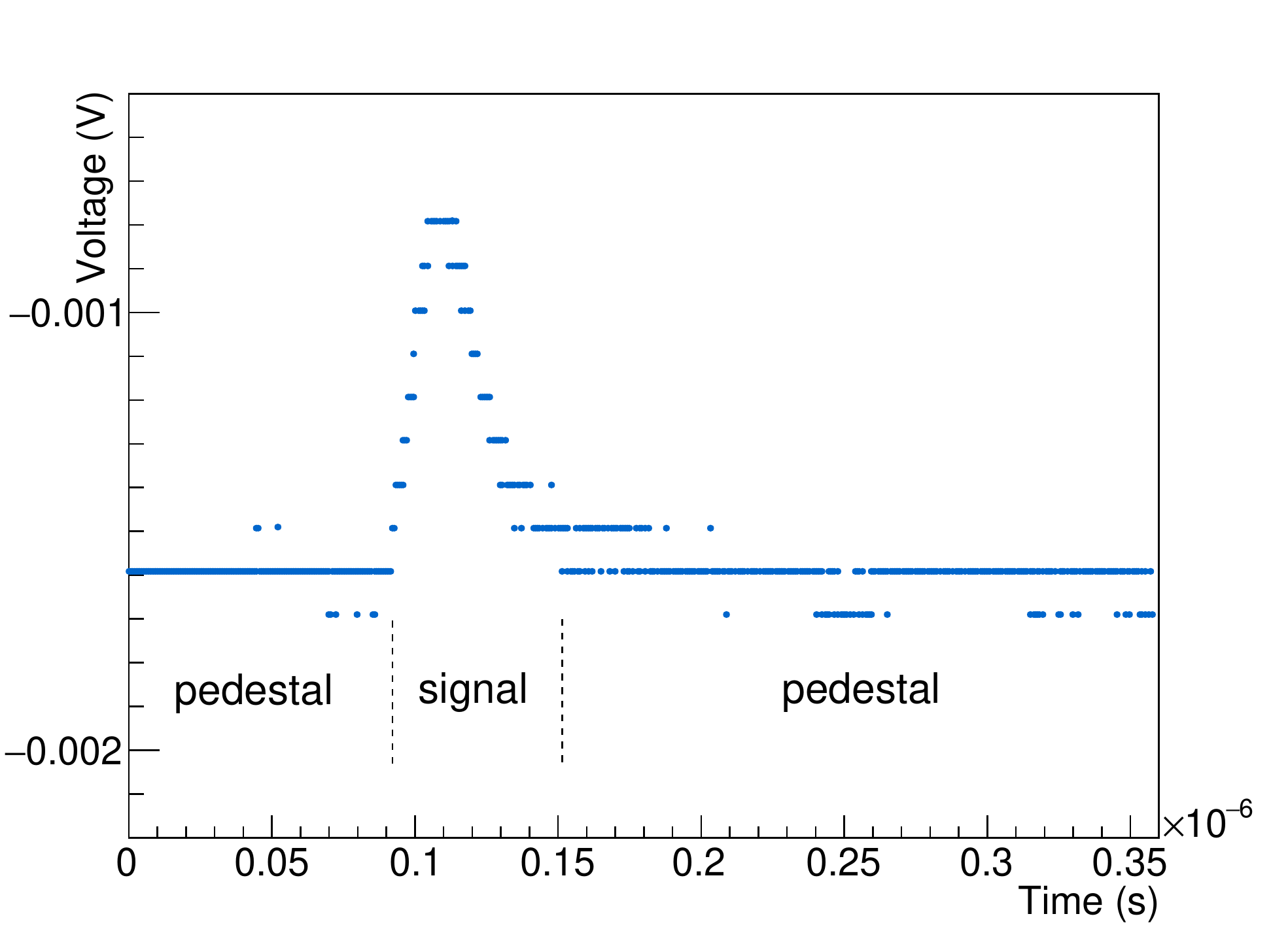}}
\caption{A typical signal pulse captured by a oscilloscope.  The step size of 0.1 mV corresponds to the accuracy of the oscilloscope.
\label{fig:One pulse signal}}
\end{center}
\end{figure}

\begin{figure}[htpb]
\begin{center}
\subfigure[]{
\label{fig:55Fe X-ray spectrum1}
\includegraphics[width=7.1cm]{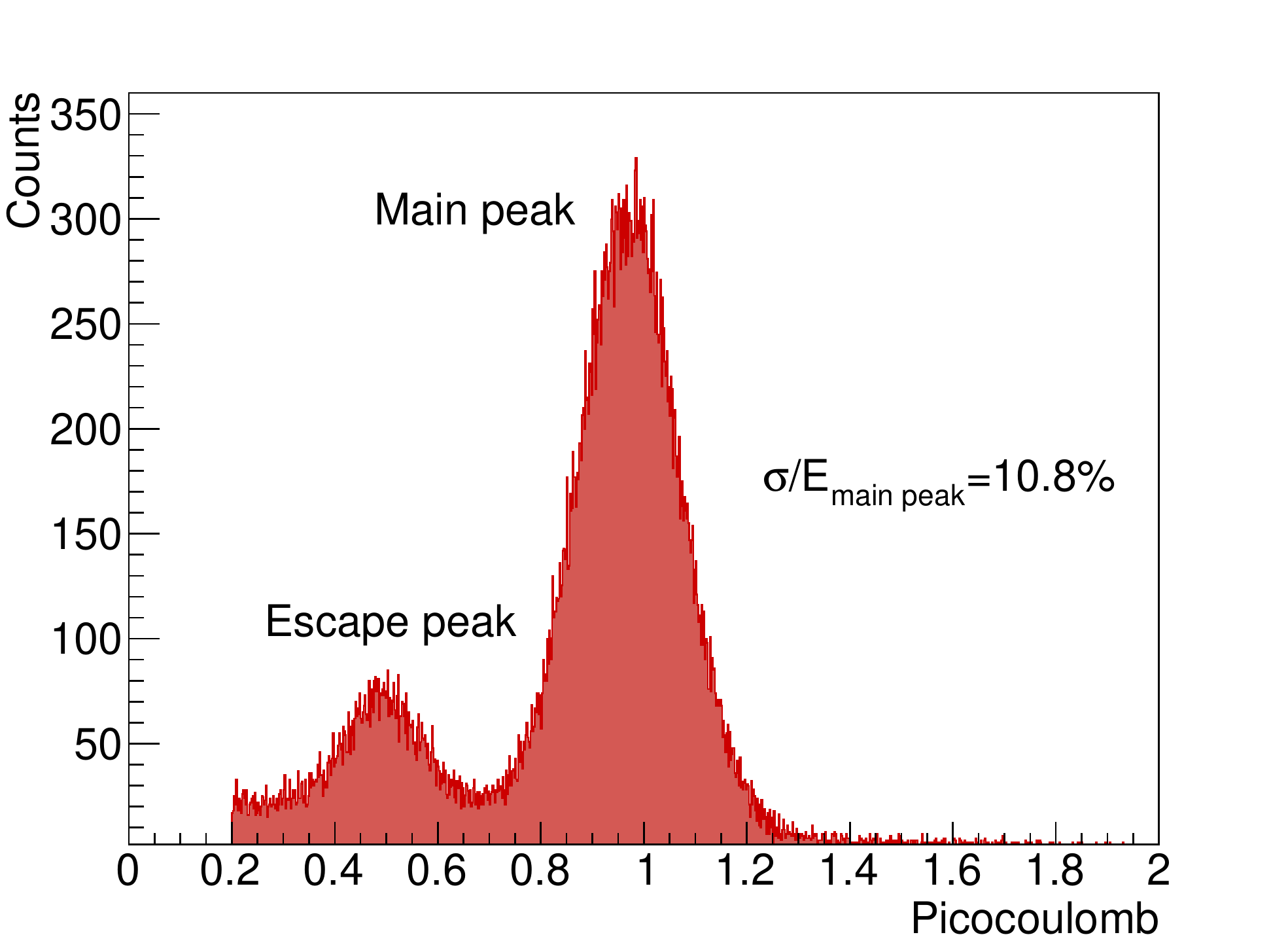}}
\\
\subfigure[]{
\label{fig:55Fe X-ray spectrum2}
\includegraphics[width=7.1cm]{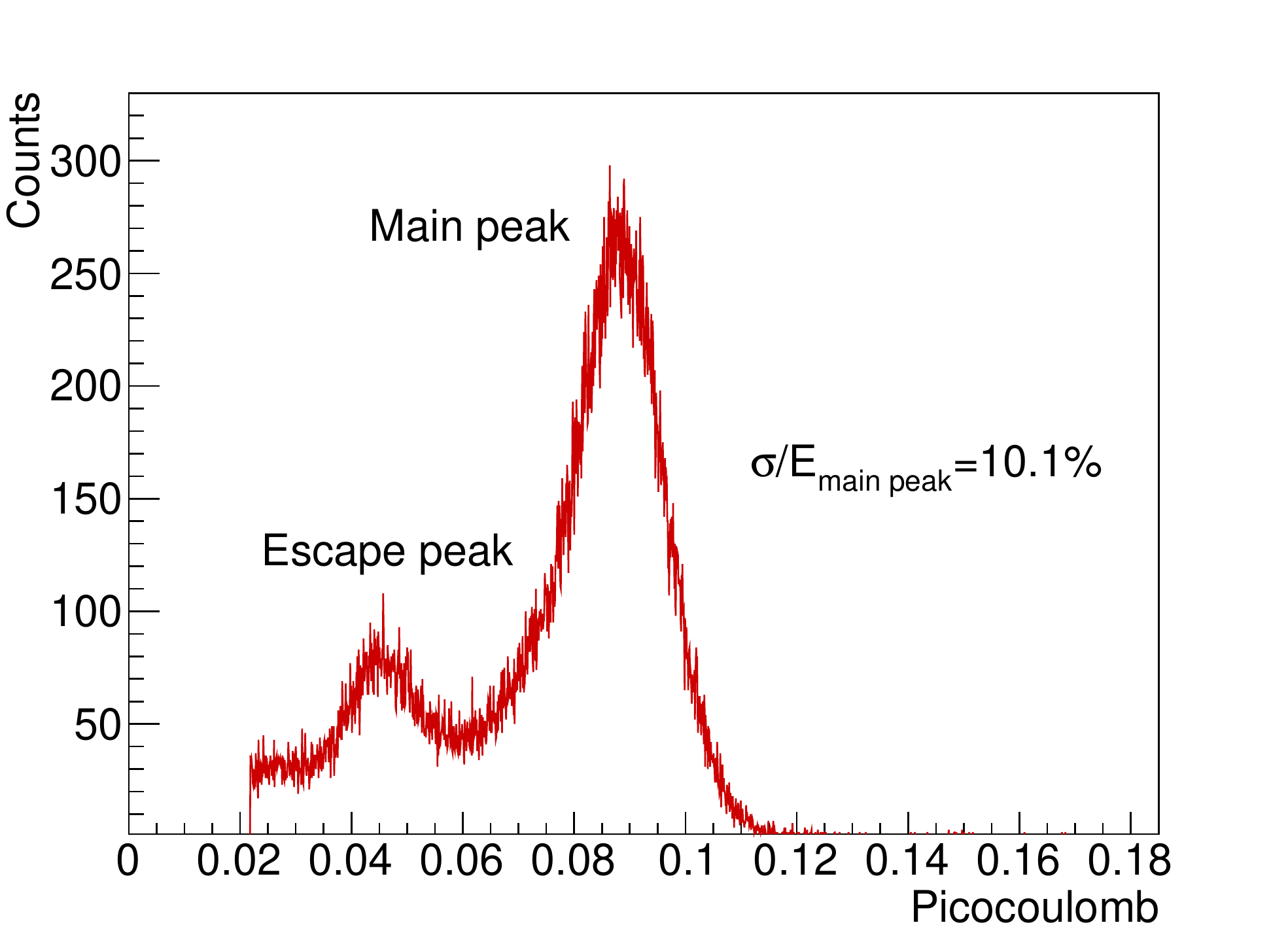}}
\caption{(a) The $^5$$^5$Fe spectrum of anode wire with $V_{\rm anode\;wires}$=1120V tested by oscilloscope. (b) The $^5$$^5$Fe spectrum of anode wire with $V_{\rm anode\;wires}$=940V obtained with MCA.
}
\label{fig:55Fe X-ray spectrum}
\end{center}
\end{figure}

\begin{figure}[htpb]
\begin{center}
\centerline{\includegraphics[width=7.3cm]{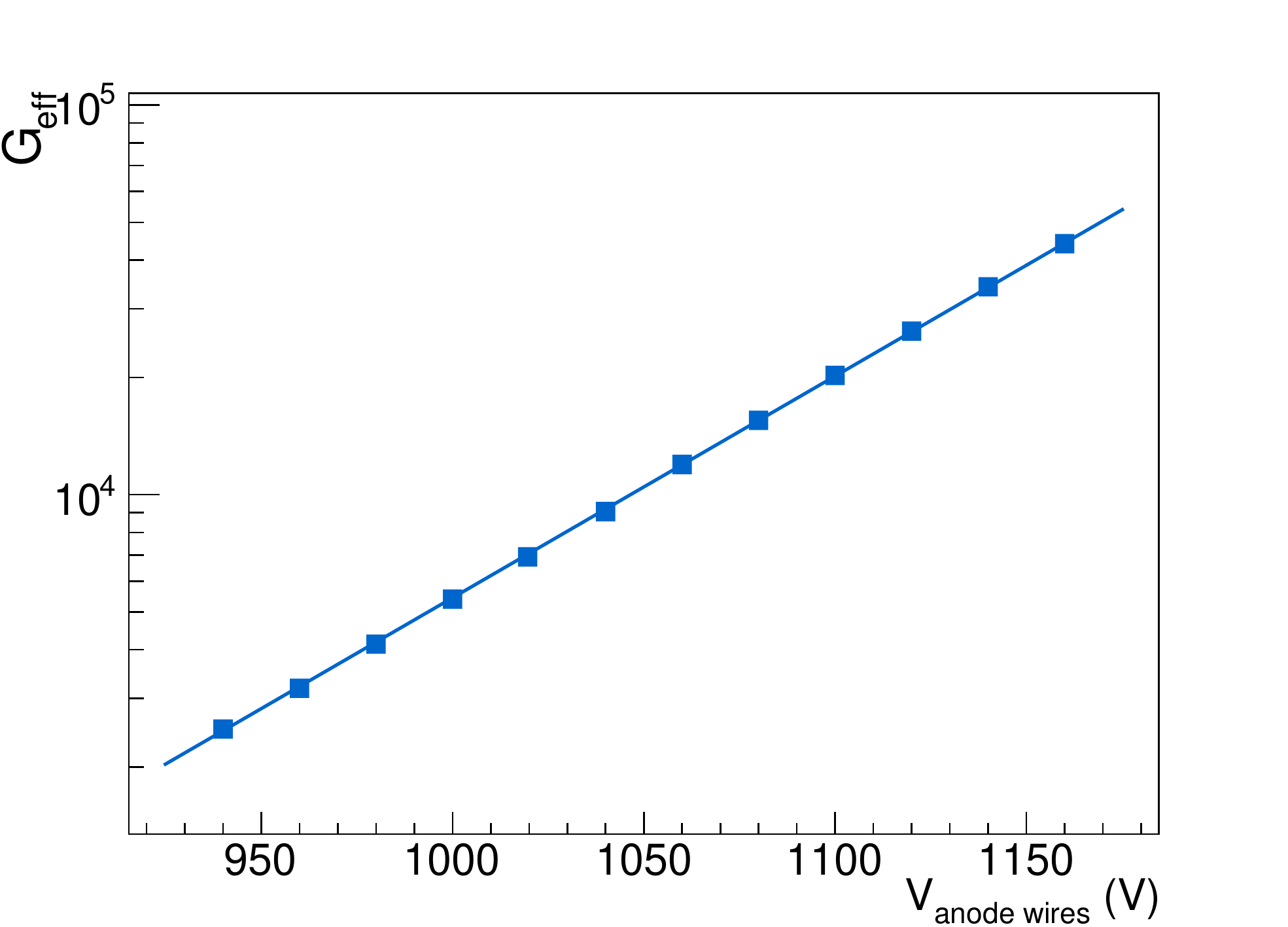}}
\caption{ Effective gain scan over anode high voltage $V_{\rm anode\;wires}$. 
\label{fig:Gain and energy resolution}}
\end{center}
\end{figure}

Figure~\ref{fig:One pulse signal} shows a typical $^5$$^5$Fe pulse signal captured by oscilloscope from one anode wire at $V_{\rm anode\;wires}$=1120V.
The average value of the data points before the rising edge is used as the pedestal $V_{\rm pedestal}$.
After pedestal substraction, the amplitude of the pulse was used to calculated the charge generated in the avalanche on anode wire.
Based on the internal resistance $P_{\rm osc}$ (50$\Omega$) and the frequency $f$ (2.5GHz) of the oscilloscope,  the charge integral can be obtained from the signal pulse.
The total amount of charge $Q_{\rm signal}$ carried by a pulse signal is calculated by:

\begin{equation}
Q_{\rm signal}=\sum_i \frac{V_{i,\rm signal}-V_{\rm pedestal}}{P_{\rm osc}}\cdot\frac{1}{f}
\label{eq:1}
\end{equation}

Figure~\ref{fig:55Fe X-ray spectrum1} shows the distribution of $Q_{\rm signal}$ for $\sim$50000 events under the same voltage set up as in Figure \ref{fig:One pulse signal}.
One main peak at 0.97pC and one escape peak at 0.48pC can be seen.
The main peak position corresponds to an effective gain of $G_{\rm eff}\sim$2.6$\times$10$^4$, and its energy resolution reaches 10.8$\%$.
Figure~\ref{fig:55Fe X-ray spectrum2} shows the $^5$$^5$Fe X-ray spectrum tested by MCA with $V_{\rm anode\;wires}$=940V (other voltages unchanged), where a more realistic  effective gain of $G_{\rm eff}\sim$2500 is achieved, with an energy resolution of 10.1$\%$.
Most of the large TPC detectors including STAR-iTPC \cite{ref:detector_07} and GEM TPC upgrade at ALICE plan to run at the gain $\sim 2500$ \cite{ref:detector_13}.

Figure~\ref{fig:Gain and energy resolution} shows the anode wire high voltage scan for the effective gain.
The gain exponentially increases with the anode high voltage as shown by the exponential fit.

\subsection{IBF ratio measurement}

Picoammeter (Keithley 6482) was used to measure the current in each part generated by gain and IBF.
The current data were acquired by the picoammeter using LabVIEW.
The background current was measured by removing the radioactive source while keeping the same working condition, which was found to be around 32pA.
To measure the IBF ratio, the currents from anode wire ($I_{\rm anode}$) and cathode/drift plane ($I_{\rm cathode}$) were first measured at each voltage set up.
The IBF ratio then can be obtained as \cite{ref:detector_12}: \footnote{There are also other definitions of IBF, i.e., $I_{\rm cathode}$/$I_{\rm anode}$ as in Ref.\cite{ref:detector_13}.} 

\begin{equation}
{IBF}=\frac{I_{\rm cathode}-{I_{\rm primary}}}{I_{\rm anode}},
\label{eq:4}
\end{equation}
where $I_{\rm primary}$ = $I_{\rm anode}$ / $G_{\rm eff}$ corresponds to the current from the primary ions by charged tracks ionization in P10 gas.

\begin{figure}[htpb]
\begin{center}
\centerline{\includegraphics[width=8.0cm]{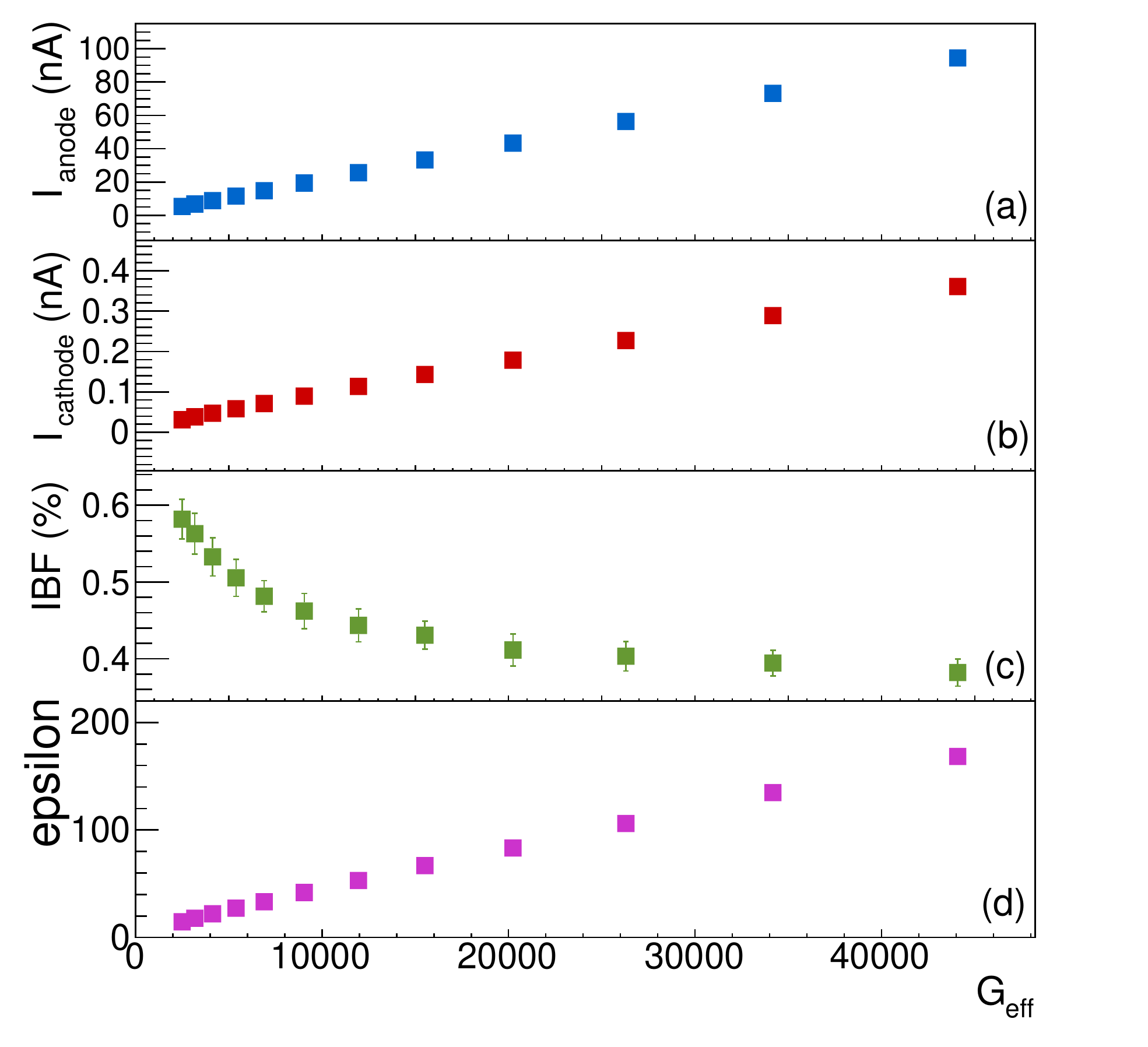}}
\caption{Experimental results of $I_{\rm anode}$, $I_{\rm cathode}$, IBF ratio and $\epsilon$ versus effective gain in staggered mode for two GEM foils. The error bars in panels (a) (b) and (d) are too small to be visible .
\label{fig:IBF measurement1}}
\end{center}
\end{figure}

Figure~\ref{fig:IBF measurement1} shows the measurement results of $I_{\rm anode}$, $I_{\rm cathode}$, IBF ratio and $\epsilon$ (=$G_{\rm eff}*$IBF) versus different effective gain (or different anode voltages) with  $E_{\rm drift}$=0.1kV/cm and $E_{\rm transfer}$=4.0kV/cm.
As we mentioned earlier in Section 2,  the staggered mode for two GEM foils in experiment was obtained by rotating 90 degree as in Figure~\ref{fig:GEM study}(c), while the full staggered mode as Figure~\ref{fig:GEM study}(b) by shifting method was realized in simulation. 
Although the rotation method is less affected by mechnical precision as the shifting method, the uncertainty caused by mounting precision can not be avoided. 
To estimate such uncertainty, multiple measurements have been performed by repeating the IBF measurements after re-mounting the GEM foils each time.  We repeated 10 times for such re-mounting and measurement procedure.  
The results with lowest IBF is used in Figure \ref{fig:IBF measurement1} to show the dependence on $G_{\rm eff}$. 

Both $I_{\rm anode}$ and $I_{\rm cathode}$ increase with the effective gain as expected.
The IBF ratio starts at about 0.58$\%$ at $G_{\rm eff}$ $\sim$2500, then decreases with increasing effective gain, which further goes down to 0.38$\%$ with $G_{\rm eff}$$>$30000.
The uncertainties on the current data are obtained from the RMS of 1000 measurements for each data point, which are about 3$\%$ and invisiable in the upper two panels.
These uncertainties are then transformed to the uncertainty for IBF ratio, which is about 5$\%$ and shown as vertical bar in Figure~\ref{fig:IBF measurement1}.

$\epsilon=G_{\rm eff}$*IBF, is defined as the number of ions flowing back to drifting region per electron from the amplification region, which usually serves as a good measure of space charge density caused by IBF.
It can be seen from Figure~\ref{fig:IBF measurement1} (c) and (d),  the IBF ratio decreases from $\sim$0.6$\%$ to below 0.4\% when $G_{\rm eff}$ increases from $\sim$2500 to $\sim$44000, and $\epsilon$ increases from $\sim$14 to $\sim$167 almost linearly.
A high $\epsilon$, consequently high space charge density should be avoided in a TPC.
Therefore, the effective gain about 2500 is a reasonable choice for a real experiment to keep a low level of space charge density.

\begin{figure}[htpb]
\begin{center}
\centerline{\includegraphics[width=8.0cm]{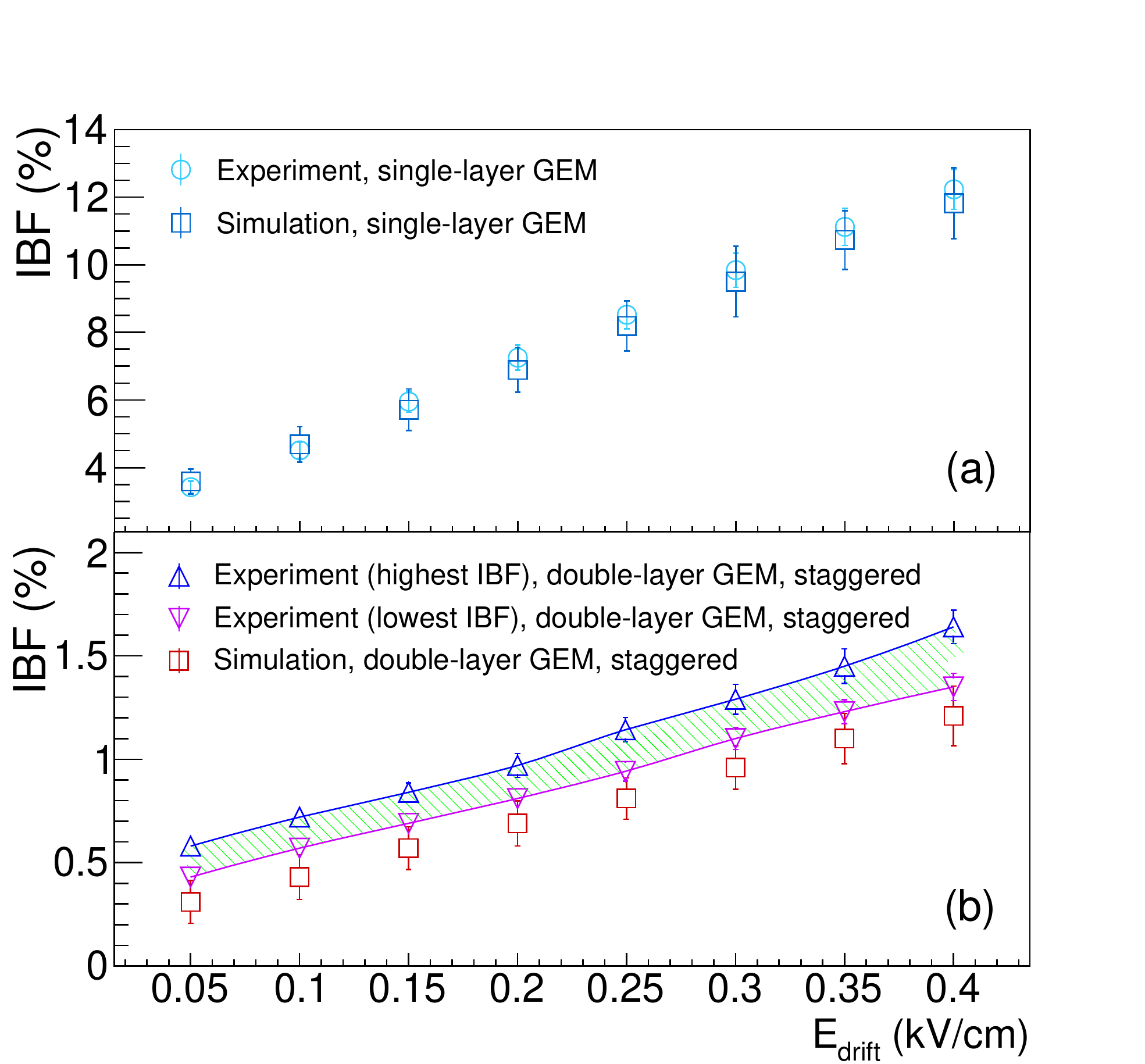}}
\caption{Comparison of measurement and simulation results for IBF ratio with a fixed gain ($G_{\rm eff}$$\sim$2500) versus $E_{\rm drift}$ (a) single GEM foil: $V_{\rm anode\;wires}$=1050V, $\Delta$$V_{\rm GEM}$=255V, $E_{\rm transfer}$=4.0kV/cm; (b) double GEM foils (staggered mode): $V_{\rm anode\;wires}$=940V, $\Delta$$V_{\rm GEM_{\rm upper}}$=$\Delta$$V_{\rm GEM_{\rm lower}}$=255V and $E_{\rm transfer}$=4.0kV/cm.
\label{fig:IBF measurement2}}
\end{center}
\end{figure}

We further checked the IBF ratio versus different $E_{\rm drift}$ with $G_{\rm eff}$$\sim$2500 for single-layer GEM mode and double-layer GEM mode.
Figure~\ref{fig:IBF measurement2} (a) shows the IBF results for single-layer GEM mode, and the measurements are in good agreement with simulation results.
The results with staggered mode of double-layer GEM are shown in Figure~\ref{fig:IBF measurement2} (b), in comparison with simulation results.
For the measurements, both the lowest and highest IBF ratios are shown from 10 measurements with rotation method for staggered mode  as mentioned above.
The band between them indicates the uncertainty on IBF related with mechnical precision when mounting the GEM foils.
In both single-layer GEM and double-layer GEM cases, IBF ratio increases linearly with increasing $E_{\rm drift}$ as expected.
However, the ion drift velocity is also increased linearly with the electric field and thus ions stay shorter in the drift volume.
As a result, the space charge density in the drift area stays the same versus $E_{\rm drift}$.

In double-layer GEM staggered mode, the IBF ratio is significantly reduced compared to the single-layer case.
The measured IBF values in double-layer case are higher than the simulation results, since the holes of two GEM foils are not 100\% staggered with rotation method in experiment while full staggering is realized in the simulation, as mentioned in Section 2.
In the staggered mode at $E_{\rm drift}$= 0.1kV/cm, the IBF can be suppressed to 0.58$\%$ $\sim$ 0.71$\%$ depending on the mechanical precision, which are reasonably low.
For example, IBF ratio about 0.7$\%$ can be reached with four layers of GEM readout set up at ALICE TPC upgrade~\cite{ref:detector_04,ref:detector_13}.

\subsection{Space charge density at EIC}

As known, the ion back flow from avalanche (IBF) and the ionization from charged particles in the drifting region will lead to space charge distortion, so the actual space charge density should be kept at a reasonable level in a real TPC detector.
A possible application for the prototype TPC chamber proposed here is the high luminosity $e+p$ collision as EIC, where the collision rate is high and thus continuous readout is a necessity.
In table \ref{tab:2}, the track density and the space charge impact to tracking in a TPC acceptance are listed for different collisions based on EIC projection and achieved luminosities at RHIC/STAR \cite{ref:Yang15,ref:detector_new02}. 
We can see that, although collision rate is high enough at EIC, but the track multiplicity is smaller than $p+p$ and significantly lower than $A+A$ collisions.
Even after considering the ions drifting back from avalanche (IBF), i.e., $\epsilon\,\sim\,14$ as obtained from our prototype TPC, the space charge situation will be similar as the $A+A$ at STAR TPC, which is acceptable. 
Therefore, the space charge density with an IBF of or below 0.7\% can probably be controlled with a reasonable way.  
The electric field distortion and thus tracking distortion due to the ions accumulated in the drifting volume can be studied with detailed simulation \cite{ref:detector_15} or even analytically by solving Langevin equation \cite{ref:detector_14, ref:detector_18,Rossegger:2011zz}.

\newcommand{\tabincell}[2]{\begin{tabular}{@{}#1@{}}#2\end{tabular}}
\begin{table*}[htbp]
\begin{center}
\begin{tabular}{|c|c|c|c|c|c|c|c|}
\hline
\tabincell{c}{Beam\\species} & \tabincell{c}{$\sqrt s$\\(GeV)} & \tabincell{c}{Peak\\Luminosity\\(cm$^{-2}s^{-1}$)} & \tabincell{c}{Cross\\section\\(cm$^{2}$)} & Nch/d$\eta$ & \tabincell{c}{Track density\\(dNch/d$\eta$ MHz)} & \tabincell{c}{Hit density\\impact hit\\finding} & \tabincell{c}{Space charge\\impact tracking}\\
\hline
e+p & 5$\times$250 & 10$^{34}$ & 10$^{-28}$ & 0.7 & 0.7 & &\\
\hline
Au+Au & 100$\times$100 & 5$\times$10$^{27}$ & 7$\times$10$^{-24}$ & 161 & 6 & Minor & \tabincell{c}{Corrected to\\good precision}\\
\hline
p+p & 100$\times$100 & 5$\times$10$^{31}$ & 3$\times$10$^{-26}$ & 2 & 3 & Minor & \tabincell{c}{Corrected to\\good precision}\\
\hline
p+p & 250$\times$250 & 1.5$\times$10$^{32}$ & 4$\times$10$^{-26}$ & 3 & 18 & \tabincell{c}{Significant\\for inner} & \tabincell{c}{Corrected to\\acceptable}\\
\hline
\end{tabular}
\caption{Particle density in TPC acceptance for different beam conditions. The achieved luminosities are based on RHIC runs up to run11 and estimated TPC performance based on STAR TPC \cite{ref:detector_new02}. Nch/d$\eta$ is the charged multiplicity per unit pseudo-rapidity.}
\label{tab:2}
\end{center}
\end{table*}

\section{Summary}

A prototype TPC chamber consisting of two layers of GEM foils and one anode wire grid has been built in order to realize continuous readout with reasonably low IBF.
Anode wires provide the main gain, while GEM foils are used to suppress the ion back-flow.
Simulations with Garfield++  are performed and a test system based on $^5$$^5$Fe has been realized to test the IBF performance.
An ion back-flow of 0.58$\%$$\sim$0.71$\%$ has been achieved with a stable gain $\sim$2500 with anode wire voltage of 940V,  $\Delta$$V_{\rm GEM}$=255V for both GEM foils, $E_{\rm transfer}$=4.0kV/cm, $E_{\rm drift}$=0.1kV/cm.
The corresponding energy resolution is about 10$\%$ with this set up.
Experimental results are in reasonably good agreement with the simulation expectations.
These studies provide useful guidance for continuous read-out TPC detector at high luminosity experiment like EIC by combining the advantages of wire grids and GEM foils.

\section*{Acknowledgments}

We would like to thank the STAR collaborators for useful discussions. The work was supported by the National Natural Science Foundation of China (No. 11520101004), and the Major State Basic Research Development Program in China (No. 2014CB845400).

\end{document}